\documentclass[a4paper,notitlepage,nofootinbib,amsmath,amssymb,prl,aps,longbibliography]{revtex4-1}
\usepackage{slashed,bbold,bm,amsmath,amssymb,mathtools,wrapfig,epsfig} 
\usepackage[usenames]{color}
\usepackage{hyperref}
\usepackage{tcolorbox}
\hypersetup{
    colorlinks=true,       
    linkcolor=red,          
    citecolor=magenta,        
    filecolor=green,      
    urlcolor=blue           
}

\newcommand{\crt}{\\[2mm]}
\newcommand{\nn}{\nonumber}
\newcommand{\beq} {\begin{equation}}
\newcommand{\eeq} {\end{equation}}
\newcommand{\beqa} {\begin{eqnarray}}
\newcommand{\eeqa} {\end{eqnarray}}

\newcommand{\mrm}[1] {{\mathrm{#1}}}

\newcommand{\bs}[1]{\boldsymbol{#1}}

\newcommand{\cf}{{\it cf.}}
\newcommand{\ie}{{\it i.e.}}

\newcommand{\eg}{{\it e.g.}}

\newcommand{\gev}{{\mrm{GeV}}}
\newcommand{\as}{\alpha_s}

\newcommand{\la}{\Lambda}

\newcommand{\vphi}{\varphi}

\newcommand{\order}[1]{${\cal O}\left(#1 \right)$}
\newcommand{\morder}[1]{{\cal O}\left(#1 \right)}
\newcommand{\eq}[1]{(\ref{#1})}
\newcommand{\fig}[1]{Fig.~\ref{#1}}
\newcommand{\lsim}{\lesssim}   

\newcommand{\inv}[1]{\frac{1}{#1}}
\newcommand{\halft}{{\textstyle \frac{1}{2}}}
\newcommand{\quart}{{\textstyle \frac{1}{4}}}
\newcommand{\sfrac}[2]{{\textstyle\frac{#1}{#2}}}
\newcommand{\intt}{{\textstyle \int}}
\newcommand{\ket}[1]{\left\vert{#1}\right\rangle}
\newcommand{\bra}[1]{\langle{#1}\vert}

\newcommand{\com}[2]{\left[{#1},{#2}\right]}
\newcommand{\comb}[2]{\big[{#1},{#2}\big]}

\newcommand{\acomb}[2]{\big\{{#1},{#2}\big\}}

\newcommand{\Asl}{{\slashed{A}}}

\newcommand{\mE}{\mathcal{E}}

\newcommand{\mH}{\mathcal{H}}

\newcommand{\mL}{\mathcal{L}}

\newcommand{\mS}{\mathcal{S}}

\newcommand{\xv}{{\bs{x}}}

\newcommand{\yv}{{\bs{y}}}
\newcommand{\zv}{{\bs{z}}}
\newcommand{\rv}{{\bs{r}}}
\newcommand{\pv}{{\bs{p}}}

\newcommand{\Av}{{\bs{A}}}

\newcommand{\Ev}{{\bs{E}}}

\newcommand{\Pv}{{\bs{P}}}

\newcommand{\gv}{\bs{\gamma}}

\newcommand{\gz}{\gamma^0}

\newcommand{\xbj}{{x_{Bj}}}

\newcommand{\rar}{\rightarrow}
\newcommand{\lar}{\leftarrow}

\newcommand{\nv}{\bs{\nabla}}

\newcommand{\rnab}{{\overset{\rar}{\nv}}\strut}
\newcommand{\lnab}{{\overset{\lar}{\nv}}\strut}

\newcommand{\alv}{{\bs{\alpha}}}

\newcommand{\Phip}{\Phi^{(P)}}

\begin{document}
\title{Hadrons as QCD Bound States\footnote{Talk at the (virtual) {\it Quark Confinement and the Hadron Spectrum 2021} conference on 2 -- 6 August 2021 in Stavanger, Norway.}}

\author{Paul Hoyer}
\email{paul.hoyer@helsinki.fi}
\affiliation{Department of Physics, POB 64, FIN-00014 University of Helsinki, Finland}
\homepage{http://www.helsinki.fi/~hoyer/}

\begin{abstract} 

Bound state perturbation theory is well established for QED atoms. Today the hyperfine splitting of Positronium is known to \order{\alpha^7\log\alpha}. Whereas standard expansions of scattering amplitudes start from free states, bound states are expanded around eigenstates of the Hamiltonian including a binding potential. The eigenstate wave functions have all powers of $\alpha$, requiring a choice in the ordering of the perturbative expansion. Temporal ($A^0=0$) gauge permits an expansion starting from valence Fock states, bound by their instantaneous gauge field. This formulation is applicable in any frame and seems promising even for hadrons in QCD. The \order{\as^0} confining potential is determined (up to a universal scale) by a homogeneous solution of Gauss' law.

\end{abstract}


\maketitle

\parindent 0cm
\parskip.2cm

\section{1. Motivations} \label{s0}

Bound states -- even QED atoms \cite{RevModPhys.57.723} -- are inherently non-perturbative. Atomic wave functions have all powers of $\alpha$ and describe exponentially suppressed processes such as tunneling. No Feynman diagram has a bound state pole. The propagators contribute inverse powers of $\alpha$ at the Bohr momentum scale $\alpha m$, thus reordering the perturbative expansion. The sum of Feynman ``ladder'' diagrams diverge at the bound state energies for arbitrarily small values of $\alpha$. Atoms are at the borderline between hard perturbative and soft classical physics, featuring elementary quanta in a classical potential. The good news is that the truly hard as well as the soft, classical regime is familiar and under theoretical control.

The omission of basic bound state methods from modern field theory (QFT) textbooks seems unwarranted \cite{Blum:2017diy}. Quantum Mechanics is traditionally introduced via the Schr\"odinger equation, which in a field theory context should be derived from the QED action. Bound states give insights into QFT that are complementary to those of scattering amplitudes.

The non-perturbative aspects of atoms and hadrons may be related, as suggested by the successful quark model classification of the hadron spectrum. The QCD coupling $\as(Q)$ is $Q$-independent in a classical gluon field. There are indications that the coupling ``freezes'' at hadronic scales, $\as(Q \lsim 1\,\gev) \simeq 0.5$ \cite{Dokshitzer:1998nz,Dokshitzer:1998qp,Deur:2016tte}. Then the strong binding of hadrons is not due to a large $\as$, and a perturbative expansion can be relevant.

\section{2. Bound state basics} \label{s1}

The accuracy of QED applied to atoms is demonstrated by the hyperfine splitting of Positronium, the mass difference $\Delta E_b$ between the ground states of Orthopositronium $(J^{PC}=1^{--})$ and Parapositronium $(0^{-+})$ \cite{Penin:2014bea,Adkins:2015wya,Adkins:2018lvj}
\begin{align} \label{e1}
\frac{\Delta E_b}{m_e} =& \frac{7}{12}\alpha^4 - \Big(\frac{8}{9}+\frac{\ln2}{2}\Big)\frac{\alpha^5}{\pi}-\frac{5}{24}\alpha^6\ln\alpha +\bigg[\frac{1367}{648}-\frac{5197}{3456}\pi^2 +\Big(\frac{221}{144}\pi^2+\inv{2}\Big)\ln2-\frac{53}{32}\zeta(3)\bigg]\frac{\alpha^6}{\pi^2} \nn\crt
&-\frac{7\alpha^7}{8\pi}\ln^2\alpha + \Big(\frac{17}{3}\ln2-\frac{217}{90}\Big)\frac{\alpha^7}{\pi}\ln\alpha +\morder{\alpha^7}
\end{align}
The $\log\alpha$ dependence arises from the quenching of infrared singularities: Soft photons decouple from the neutral atom when their wavelength is larger than its Bohr radius $r = 2/\alpha m_e$. 

Despite impressive achievements such as \eq{e1} the principles of bound state perturbation theory are not formulated at a level similar to the perturbative $S$-matrix in the Interaction Picture. The Poincar\'e invariance of interacting, spatially extended states is non-trivial.
To illustrate, atoms in motion are often depicted as ellipses due to Lorentz contraction, with no discussion of how the contraction is realized in QED. The effective binding energy $\Delta E(P)$ of a Positronium atom of mass $M=2m_e+E_b$, where $E_b = -\quart \alpha^2 m_e+{\cal O}(\alpha^4)$, is inversely proportional to its momentum $P$ in the limit of relativistic motion, $P \gg 2m_e$:
\begin{align} \label{e2}
\Delta E(P) \equiv \sqrt{P^2+(2m_e+E_b)^2} - \sqrt{P^2+4m_e^2} = \frac{2m_e E_b}{P} +\morder{\alpha^4}
\end{align}
The Coulomb energy $V(\rv) = -\alpha/|\rv|$ is independent of $P$ for the uncontracted orthogonal separations, $\rv\cdot\Pv=0$. This seems to contradict the $P$-dependent energy difference \eq{e2}. An explanation was given in \cite{Jarvinen:2004pi} using the Bethe-Salpeter (B-S) equation \cite{Salpeter:1951sz,Itzykson:1980rh}. For $P>0$ transverse photon exchange contributes at leading \order{\alpha^2} to the  binding energy. This cancels the Coulomb energy (see Eq. (33) of \cite{Jarvinen:2004pi}), leaving an interaction that ensures the correct dependence of the Positronium energy on its momentum, $E(P)= \sqrt{M^2+P^2}$. The valence Fock state wave function $\ket{e^+e^-}$ contracts as expected, whereas $\ket{e^+e^-\gamma}$ behaves differently.

Feynman diagrams arise in an expansion around free states, which lack overlap with bound states. Consequently, no Feynman diagram for $e^+e^- \to e^+e^-$ has a pole at the Positronium energies. Bound state binding may be recovered by expanding the electron propagator in a photon field $A^\mu$ in a geometric series with free propagators,
\begin{align} \label{e3}
\frac{i}{i\slashed{\partial}-m-e\Asl} = \frac{i}{i\slashed{\partial}-m} - \frac{i}{i\slashed{\partial}-m}ie\Asl \frac{i}{i\slashed{\partial}-m} + \ldots
\end{align}
In the non-relativistic limit Positronium dynamics is equivalent to that of the $e^-$ in a fixed $A^0$ field. This allows to obtain the Schr\"odinger equation through a sum of Feynman ladder diagrams analogous to \eq{e3} (see, \eg, Section III of \cite{Hoyer:2021adf}). The Positronium poles of $e^+e^- \to e^+e^-$ arise through the divergence of the sum of ladder diagrams.

Relativistic bound state dynamics cannot be reduced to scattering in an external field. Higher order corrections to Positronia were first evaluated \cite{Murota:1988hr} using the Bethe-Salpeter equation. It is based on the Dyson-Schwinger (D-S) identity for a $2 \to 2$ Green function $G_T$, whose external propagators are truncated,
\begin{align} \label{e4}
G_T = K + G_T\,S\,K
\end{align}
Here $S$ is a two-particle propagator and $K$ a two-particle irreducible kernel. Any Feynman diagram contributes equally to the two-sides of \eq{e4}. The D-S equation is thus formally exact when $S$ and $K$ are expanded to all orders in $\alpha$. The B-S equation for bound state wave functions $\Phi_T$ is given by the residues of the poles in $G_T$,
\begin{align} \label{e4b}
\Phi_T = \Phi_T\,S\,K
\end{align}

Being based on Feynman diagrams the B-S equation is Poincar\'e covariant, but notoriously difficult to solve. There are retardation effects even at lowest order, with free Dirac propagators in $S$ and a single photon exchange kernel $K$: The $A^\mu$ field at $e^-(t,\xv)$ depends on the distribution of $e^+(t-t',\xv')$ at earlier times, due to the propagation time $t'$ of transverse photons. No analytic solution of the B-S equation is known.

The rhs. of the D-S equation \eq{e4} involves a double perturbative expansion in $S$ and $K$, whereas $G_T$ has a unique expansion in $\alpha$ \cite{Caswell:1978mt,Lepage:1978hz}. This allows to fix $S$ and let the expansion of $K$ be determined by that of $G_T$. The B-S equation can thus at lowest order in the rest frame be reduced to the Schr\"odinger equation, and the higher order corrections systematically determined. More intuitively: There are many equivalent expansions in $\alpha$ when the first approximation of the wave function already has all powers of $\alpha$. All reorderings must give the same result for physically measurable quantities such as the binding energies in \eq{e1}.

The realization that the B-S equation is not unique led to Non-Relativistic QED \cite{Caswell:1985ui,Kinoshita:1998jfa}. NRQED expands in powers of $|\pv|/m_e$, the velocity of the $e^\pm$ in the rest frame. This reordering of the original B-S expansion is motivated by the nearly non-relativistic motion of atomic electrons. The expansion is uniquely defined when the Schr\"odinger equation is chosen as starting point. The higher order terms in \eq{e1} were derived using NRQED.

\section{3. The instantaneity of gauge interactions} \label{s2}

In the Schr\"odinger approximation Positronium is described by its valence $\ket{e^+e^-}$ Fock state, which is an eigenstate of the QED Hamiltonian with the instantaneous Coulomb potential. The tranverse photon interaction terms in $\mH_{QED}$ create Fock states such as $\ket{e^+e^-\gamma}$, which contribute to the binding energy at higher orders in $\alpha$. This suggests \cite{Hoyer:2021adf} a ``Bound Fock expansion'', where the constituents of each Fock state propagate in their instantaneous gauge field. Transitions between Fock states are given by the transverse photon interactions of $\mH_{QED}$, and are suppressed by the coupling $e$. With all Fock states and the full Hamiltonian taken into account the method is formally exact and valid in any frame.

It is sometimes held that the concept of an instantaneous potential is valid only in a non-relativistic limit. However, gauge theories have instantaneous interactions even with relativistic dynamics. This seems surprising since the action is local and no constituent moves faster than light. The exception arises because the gauge can be fixed non-locally. Thus all instantaneous interactions are gauge dependent.

The QED action
\begin{align} \label{e5}
\mS_{QED} &= \int d^4x\big[-\quart F_{\mu\nu}F^{\mu\nu}+\bar\psi(i\slashed{\partial}-m-e\slashed{A})\psi\big]
& F_{\mu\nu} = \partial_\mu A_\nu-\partial_\nu A_\mu
\end{align}
has no term involving $\partial_0 A^0$ since $F_{00}=0$. Similarly, $F_{ij}$ gives no space derivative of the longitudinal gauge field, which is defined by a scalar potential, $\Av_L^j = \partial_j\, \phi$. Hence $A^0$ and $\Av_L$ do not propagate: Their values are fixed by the gauge. 

Feynman gauge is convenient for scattering amplitudes, as the gauge fixing term $\mS_{GF}=-\halft\int d^4x\,(\partial_\mu A^\mu)^2$ preserves full Poincar\'e invariance. $\mS_{GF}$ adds the ``missing'' derivatives
$\partial_0 A^0$ and $\nv\cdot\Av_L$ to the action, allowing the gauge dependent fields to propagate similarly to the gauge invariant ones. For bound states, on the other hand, instantaneous interactions are welcome due to their lack of retardation. It is then advantageous to use Coulomb gauge ($\nv\cdot \Av_L=0$) or temporal gauge ($A^0=0$), which non-locally fix $\Av_L$ and $A^0$, respectively.

The Hamiltonian generates evolution in time, leaving only space rotations and translations as explicit (``kinetic'') symmetries. Time translations and boosts involve interactions and are thus ``dynamic'' symmetries. Both the Coulomb and temporal gauge conditions preserve the kinetic symmetries of the Hamiltonian.

The Hamiltonian approach defines a conjugate field $\pi_\alpha$ for each field $\vphi_\alpha$ in the action, and imposes canonical commutation relations between them,
\begin{align} \label{e6}
\pi_\alpha(t,\xv) \equiv \frac{\delta\mS(\vphi,\partial\vphi)}{\delta[\partial_0\vphi_\alpha(t,\xv)]}
\hspace{2cm}
\com{\vphi_\alpha(t,\xv)}{\pi_\beta(t,\yv)}_\pm = i\delta_{\alpha\beta} \delta^3(\xv-\yv)
\end{align}
In QED the conjugate field of $A^0$ vanishes. Quantization in Coulomb gauge can be realized by means of constraints \cite{Weinberg:1995mt}, which are cumbersome especially in QCD \cite{Christ:1980ku}. Temporal gauge eliminates $A^0$ and thus avoids the problem with its lack of conjugate field. The canonical commutation relations can then be imposed for all space components $\Av$, with the longitudinal electric field $\Ev_L=-\partial_0 \Av_L$ being conjugate to $\Av_L$. 

The classical nature of the instantaneous $\Ev_L$ field is apparent in temporal gauge. The gauge condition $A^0=0$ needs to be supplemented with a condition on the remaining gauge degrees of freedom \cite{Willemsen:1977fr,Bjorken:1979hv,Leibbrandt:1987qv}. Time independent gauge transformations which preserve $A^0=0$ are generated by Gauss' operator,
\begin{align} \label{e7}
G(x) \equiv \frac{\delta\mS_{QED}}{\delta{A^0(x)}} = \partial_i E^{i}_L(x)-e\psi^\dag\psi(x)
\end{align}
Since $A^0$ is fixed $G(x)=0$ is not an operator equation of motion, but a condition which is imposed on physical states, $G(x)\ket{phys}=0$ (similarly as $\Av_L\ket{phys}=0$ in the Gupta-Bleuler formalism). This ensures that the $\ket{phys}$ states are invariant under time independent gauge transformations and determines the value of $\Ev_L$,
\begin{align} \label{e8}
\Ev_L(t,\xv)\ket{phys} &= -\nv_x \int d\yv\, \frac{e}{4\pi|\xv-\yv|}\,\psi^\dag\psi(t,\yv)\ket{phys}
\end{align}
This is not an operator constraint, \eg, it should be understood that $\psi^\dag\psi(t,\yv)\ket{0}=0$.
The electric field $\Ev_L$ is classical, generated by the charged constituents and contributes an instantaneous (``$V$'') potential to the QED Hamiltonian in temporal gauge,
\begin{align} \label{e9}
\mH_{QED} &= \int d\xv\,\big[E^i\partial_0 A_i+i\psi^\dag\partial_0\psi-\mL_{QED}\big] = \mH_{QED}^{(V)} + \mH_{QED}^{(KI)} \nn\crt
\mH_{QED}^{(V)} &= \int d\xv\,\halft \Ev^2_L =\halft\int d\xv d\yv\, \frac{\alpha}{|\xv-\yv|}\big[\psi^\dag\psi(t,\xv)\big]\big[\psi^\dag\psi(t,\yv)\big] \nn\crt
\mH_{QED}^{(KI)}&= \int d\xv\,\big[\halft \Ev^2_T +\quart F^{ij}F^{ij}+\psi^\dag\big[(-\alv\cdot(i\nv+e\Av)+m\gz)\big]\,\psi\big] 
\end{align} 
where $\alv \equiv \gz\gv$. The kinetic energies of the electron and photon fields, as well as their $-e\int d\xv\,\psi^\dag\alv\cdot\Av\,\psi$ interaction term, are in the ``$KI$'' part of the Hamiltonian.

\section{4. Application to Positronium} \label{s3}

The valence $\ket{e^+e^-}$ Fock state of Positronium is defined in terms of the electron field $\psi(x)$ as
\begin{align} \label{e10}
\ket{e^-e^+;\Pv} \equiv \int d\xv_1 d\xv_2\,\bar\psi(\xv_1)e^{i\Pv\cdot(\xv_1+\xv_2)/2}\,\Phi^{(\Pv)}(\xv_1-\xv_2)\psi(\xv_2)\ket{0}
\end{align}
where  $t_1=t_2=0$. The perturbative vacuum $\ket{0}$ projects on the creation operators $b^\dag,d^\dag$ in $\bar\psi$ and $\psi$, respectively. The ($c$-number) wave function $\Phi^{(\Pv)}$ has $4\times 4$ Dirac components and depends on the bound state momentum $\Pv$ and on the quantum numbers of the Positronium.

The action of $\mH_{QED}^{(V)}$ \eq{e9} on the state \eq{e10} follows from $\comb{\psi^\dag\psi(\xv)}{\bar\psi_\alpha(\xv_1)}=\bar\psi_\alpha(\xv_1)\delta(\xv-\xv_1)$ and $\comb{\psi^\dag\psi(\yv)}{\psi_\beta(\xv_2)}=-\psi_\beta(\xv_2)\delta(\yv-\xv_2)$. Together with the $\xv\leftrightarrow\yv$ contribution this gives the Coulomb energy (the infinite ``self-energies'' with $\xv=\yv=\xv_1$ and $\xv=\yv=\xv_2$ are subtracted)
\begin{align} \label{e11}
\mH_{QED}^{(V)}\,\bar\psi_\alpha(\xv_1)\psi_\beta(\xv_2)\ket{0} = -\frac{\alpha}{|\xv_1-\xv_2|}\,\bar\psi_\alpha(\xv_1)\psi_\beta(\xv_2)\ket{0}
\end{align}
In the rest frame ($\Pv=0$) only the kinetic energy of the electrons in $\mH_{QED}^{(KI)}$ contributes at leading order, since the radiation of a transverse photon is suppressed by the $e^\pm$ Bohr momentum $\alpha m/2$. The eigenstate condition gives the bound state equation (BSE),
\begin{align} \label{e12}
\mH_{QED}\,\ket{e^-e^+;\Pv=0} &= M\,\ket{e^-e^+;\Pv=0} \nn\crt
\big(i\alv\cdot\rnab+m\gz\big)\Phi^{(0)}(\xv)+\Phi^{(0)}(\xv)&\big(i\alv\cdot\lnab-m\gz\big) = \big[M- V(\xv)\big]\Phi^{(0)}(\xv)
\end{align}
where $V(\xv)=-\alpha/|\xv|$. This BSE reduces to the Schr\"odinger equation in the non-relativistic limit, which may be taken at \order{\alpha^2} in the binding energy $E_b = M-2m$.

For Positronium in motion $(P >0)$ the $e^{\pm}$ have momenta $\halft \Pv+\morder{\alpha m}$, so the $ee\gamma$ vertices are no longer suppressed by $\alpha$. Including the $\ket{ee\gamma}$ Fock state \cite{Hoyer:2021adf} gives agreement with the B-S analysis \cite{Jarvinen:2004pi}: The $\ket{e^+e^-}$ wave function Lorentz contracts and the bound state energy $E=\sqrt{M^2+P^2}$ as required by Poincar\'e invariance. 

Transverse photon exchange as well as annihilation $e^+e^- \to \gamma^* \to e^+e^-$ give the Positronium hyperfine splitting \eq{e1} at \order{\alpha^4}. First checks of bound state scattering were also made for form factors and deep inelastic scattering \cite{Hoyer:2021adf}. The bound Fock expansion outlined here may seem cumbersome, but should remove the ``art'' (noted on p. 493 of \cite{Itzykson:1980rh}) from bound state calculations. Calculations can be streamlined when the foundation is solid.

\section{5. Application to hadrons} \label{s4}

The properties of charmonium and bottomonium states are successfully described using the Schr\"odinger equation with the ``Cornell potential'' \cite{Eichten:1979ms,Eichten:2007qx},
\begin{align} \label{e13}
V(r) = V'r-\frac{4}{3}\frac{\as}{r} \ \ \ \text{with}\ \ V' \simeq 0.18\ \text{GeV}^2, \ \ \as \simeq 0.39
\end{align}
The confining potential $V'r$ in \eq{e13} involves a scale $V'$ which is not present in the QCD action, and thus also not in Gauss' operator \eq{e7}. The only way this scale can appear in a classical solution of Gauss' law is through a boundary condition. Analogous scenarios have been discussed before, notably in the context of the MIT Bag Model \cite{Chodos:1974je}. Can a non-trivial boundary condition arise in the bound state approach discussed above, applied to QCD in temporal gauge? 

Gauss' operator in QCD is
\begin{align} \label{e14}
G_a(x) \equiv \frac{\delta\mS_{QCD}}{\delta{A_a^0(x)}} = \partial_i E_a^{i}(x)+g f_{abc}A_b^i E_c^i-g\psi^\dag T^a\psi(x)
\end{align}
As in QED, the color electric field $\Ev_L^a$ is constrained by $G_a(x)\ket{phys}=0$ \cite{Willemsen:1977fr,Bjorken:1979hv,Leibbrandt:1987qv},
\begin{align} \label{e15}
\nv\cdot \Ev_{L}^{a}(\xv)\ket{phys} = g\big[- f_{abc}A_b^i E_c^i+\psi^\dag T^a\psi(\xv)\big]\ket{phys}
\equiv g\, \mE_a(\xv)\ket{phys}
\end{align}
In QED the $\bar\psi(\xv_1)\psi(\xv_2)\ket{0}$ component of the state \eq{e10} generates a dipole electric field, giving rise to the Coulomb potential $-\alpha/|\xv_1-\xv_2|$ in \eq{e11}. The analogous color potential $-\sfrac{4}{3}\as/r$ in \eq{e13} arises from the color $A$ quark components $\ket{q_A(\xv_1)\bar q_A(\xv_2)}$ of a meson state. However, a color singlet meson cannot generate a classical color octet electric field $E_{L}^{a}(\xv)$ {\it at any point $\xv$}. Color symmetry ensures that the electric field cancels in the sum over quark colors. 

The QED solution \eq{e8} for $\Ev_L$ was obtained with the obvious boundary condition $\lim_{|\xv|\to\infty}\Ev_L(\xv) =0$. The situation is less obvious in QCD, where the color electric field in any case vanishes at all $\xv$ for color singlets. Confinement motivates to consider alternative boundary conditions, \ie, homogeneous solutions of \eq{e15}.

A homogenous solution should maintain the kinetic symmetries of space rotations and translations. The only possibility appears to be 
\begin{align} \label{e16}
\Ev^a_{H}(\xv)\ket{phys} &= -\kappa\,\nv_x \int d\yv \,\xv\cdot\yv\,\mE_a(\yv) \ket{phys}
\end{align}
where $\mE_a(\yv)$ is defined in \eq{e15}. $\Ev^a_{H}$ trivially satisfies $\nv_x\cdot \Ev_{H}^{a}=0$ when the normalization $\kappa$ is independent of $\xv$. Adding the homogeneous $\Ev^a_{H}$ to a particular solution of \eq{e15} (analogous to \eq{e8} in QED) defines the instantaneous interaction. After partial integration,
\begin{align} \label{e17}
\mH_{QCD}^{(V)} &\equiv \halft\int d\xv\,(\Ev_{a,L})^2 = \int d\yv d\zv\,\Big\{\,\yv\cdot\zv \Big[\halft\kappa^2\intt d\xv + g\kappa\Big] + \halft \frac{\as}{|\yv-\zv|}\Big\}\,\mE_a(\yv)\mE_a(\zv)
\end{align} 
The term of \order{\kappa^2} is $\propto\hspace{-1mm} \int d\xv$ (the volume of space) because the homogeneous solution generates an $\xv$-independent field energy density. This (infinite) contribution is irrelevant provided it is universal, \ie, the same for all components of all bound states. This determines the normalization $\kappa$ in \eq{e16} for each state $\ket{phys}$, up to a common scale $\la$.

Applying $\mH_{QCD}^{(V)}$ to a meson Fock component $\ket{q\bar q} \equiv \sum_A\bar\psi_A^\alpha(\xv_1)\psi_A^\beta(\xv_2)\ket{0}$ gives
\begin{align} \label{e18}
\mH_{QCD}^{(V)} \ket{q\bar q} = \Big[\big(\halft\kappa^2\intt d\xv + g\kappa\big)C_F\,(\xv_1-\xv_2)^2-C_F\frac{\as}{|\xv_1-\xv_2|}\Big]\ket{q\bar q}
\end{align}
where $C_F = 4/3$ for QCD.
For the coefficent of $\kappa^2$ to be independent of $\xv_1,\,\xv_2$ we must have $\kappa \propto 1/|\xv_1-\xv_2|$. A universal scale $\la$ may be defined by choosing
\begin{align} \label{e19}
\kappa_{q\bar q} = \frac{\la^2}{gC_F}\frac{1}{|\xv_1-\xv_2|}
\end{align}
The potential energy defined by $\mH_{QCD}^{(V)} \ket{q\bar q}=V_{q\bar q}(\xv_1-\xv_2)\ket{q\bar q}$ is then, apart from the universal contribution of \order{\kappa^2},
\begin{align} \label{e20}
V_{q\bar q}(\xv_1-\xv_2) = \la^2|\xv_1-\xv_2|-C_F\frac{\as}{|\xv_1-\xv_2|}
\end{align}
Encouragingly, this agrees with the phenomenological Cornell potential \eq{e13}. Note that the linear confinement was not assumed, it is a consequence of the required form \eq{e16} of the homogeneous solution of Gauss' constraint. The instantaneous potential \eq{e20} holds for quarks of any mass and momentum. Applying $\mH_{QCD}$ to a meson valence state analogous to \eq{e10} and neglecting higher Fock states gives the bound state equation \eq{e12} in the rest frame, with the potential \eq{e20}. 

The Hamiltonian $\mH_{QCD}^{(V)}$ \eq{e18} determines the instantaneous potential of any state with quarks and gluons. For a baryon component $\ket{qqq}=\sum_{A,B,C}\epsilon_{ABC}\,\psi_A^\dag(\xv_1)\psi_B^\dag(\xv_2)\psi_C^\dag(\xv_3)\ket{0}$,
\begin{align} \label{e21}
V_{qqq}(\xv_1,\xv_2,\xv_3) &=\la^2\,d_{qqq}(\xv_1,\xv_2,\xv_3)-\frac{2}{3}\,\as\Big(\inv{|\xv_1-\xv_2|}+\inv{|\xv_2-\xv_3|}+\inv{|\xv_3-\xv_1|}\Big) \nn\crt
d_{qqq}(\xv_1,\xv_2,\xv_3) &= \inv{\sqrt{2}}\sqrt{(\xv_1-\xv_2)^2+(\xv_2-\xv_3)^2+(\xv_3-\xv_1)^2}
\end{align}
When two of the quarks in the baryon are close to each other the potential should (for $N_c=3$) reduce to that of the color $3\otimes\bar 3$ meson potential $V_{q\bar q}$ \eq{e20}. Setting $\xv_2=\xv_3$ in $V_{qqq}$ (and subtracting the infinite Coulomb energy) indeed gives
\begin{align} \label{e22}
V_{qqq}(\xv_1,\xv_2,\xv_2) = \la^2 |\xv_1-\xv_2|-\frac{4}{3}\frac{\as}{|\xv_1-\xv_2|} = V_{q\bar q}(\xv_1,\xv_2)
\end{align}
Corresponding potentials are obtained for color singlet $\ket{q\bar qg}$ and $\ket{gg}$ Fock states \cite{Hoyer:2021adf}.

The above potentials depend only on the instantaneous positions $\xv_i$ of the charges. They are independent of the bound state wave function and the CM momentum $\Pv$. In a general frame, for a $q\bar q$ state of the form \eq{e10}, the bound state equation is
\begin{align} \label{e23}
i\nv\cdot\acomb{\alv}{\Phip(\xv)}-\halft \Pv\cdot \comb{\alv}{\Phip(\xv)}+m\comb{\gz}{\Phip(\xv)} &= \big[E-V(\xv)\big]\Phip(\xv)
\end{align}
where $E=\sqrt{M^2+\Pv^2}$ and the potential $V(\xv)$ is as in \eq{e20}. Like in QED, Lorentz covariance requires for $P>0$ to take into account also transverse gluon exchange, \ie, the \order{g} $\ket{q\bar qg}$ Fock state created by $\mH_{QCD}^{(KI)}$ (\cf\ \eq{e9} for QED). However, at \order{\as^0} the $\ket{q\bar q}$  state must satisfy covariance on its own, as any contribution from higher Fock states is of higher order in $\as$. Remarkably, a linear $q\bar q$ potential is indeed boost covariant by itself. The $\Pv$-dependence of the wave function is not given by standard Lorentz contraction, however.

More generally, the ${\cal O}(\as^0)$ meson states are fully determined by the $\ket{q\bar q}$ Fock states and the BSE \eq{e23} with $V(\xv)=\la^2|\xv|$. Like for the Dirac equation the wave function has negative energy components, which reflect $q\bar q$ pairs due to time ordering (``$Z$''-diagrams). These pairs have features similar to sea quarks, contributing to deep inelastic scattering at small $\xbj$. The ${\cal O}(\as^0)$ states have no (transverse) gluon constituents -- they arise at \order{g} via radiative effects. However, we may consider states with valence gluons at \order{\as^0}, such as hybrid $\ket{q\bar qg}$ or glueball $\ket{gg}$ states.

\begin{figure}[h]
\centering
\includegraphics[width=10cm]{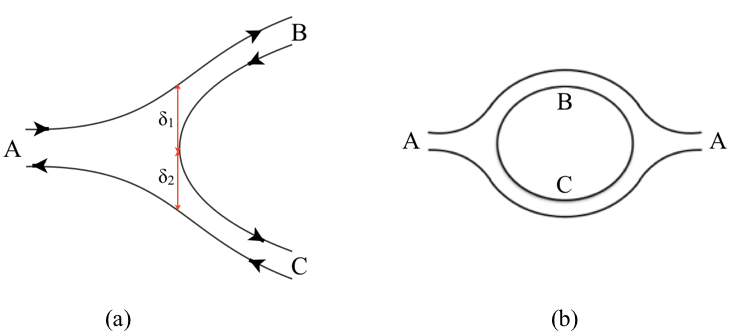}
\caption{(a) The overlap $\bra{B,C}A\rangle$, ``string breaking'', is determined by the wave functions of hadrons $A,\,B$ and $C$. The linear potential is nearly the same before and after the split, $V({\bs\delta}_1+{\bs\delta}_2) \simeq V({\bs\delta}_1)+V({\bs\delta}_2)$. (b) Hadron loop correction through the creation and annihilation of a $q\bar q$ pair.}
\label{fig-1}
\end{figure}

For small quark masses $m$ the BSE \eq{e23} at \order{\as^0} gives $\ket{q\bar q}$ states which lie on approximately linear Regge trajectories, with parallel daughter trajectories. There are no states with quantum numbers that would be exotic in the quark model. Chiral symmetry is exact for $m=0$. There are $P^\mu=0$ solutions whose mixing with the vacuum does not violate Poincar\'e invariance. This may allow to realize a spontaneous breaking of chiral symmetry.

For quark separations $\xv$ such that $V(\xv) =\la^2|\xv| \ll E$ the BSE \eq{e23} describes free quarks. This gives rise to features of the parton model and duality. ``String breaking'' at large potentials $V(\xv)$ occur through an overlap of the hadron state with two or more hadrons (\fig{fig-1}), as required for hadronization.
These and other aspects require further study.

\vspace{.5cm}

\noindent{\bf Acknowledgements.} I thank the organizers of the (virtual) {\it Quark Confinement and the Hadron Spectrum 2021} conference in Stavanger, Norway for the opportunity to present this talk. I enjoy the privileges of Professor Emeritus at the Physics Department of Helsinki University. Travel grants from the Magnus Ehrnrooth Foundation have allowed me to maintain contacts and discuss my research with colleagues.

\bibliography{210827_refs}
\end{document}